\documentclass[aps,amssymb,article,pra,twocolumn,showpacs,showkeys,floatfix]{revtex4}

\usepackage{epsf}
\usepackage{amsmath,bm}
\usepackage{graphicx}
  \begin{document}
         
  \title{The quantum state of graphene}

  \author{Boris I. Ivlev}

  \affiliation{Instituto de F\'{\i}sica, Universidad Aut\'onoma de San Luis Potos\'{\i},\\ 
  San Luis Potos\'{\i}, 78000 Mexico}

  \begin{abstract}
  
  A stationary solution of quantum mechanical wave equation is the superposition of eigenfunctions. Each of them corresponds to a vector in the Hilbert space. In a graphene
  sample one can choose expansion coefficients to get that series convergent solely within the certain circle in the two-dimensional space. Outside this circle the analytic 
  continuation is required in the form of a different series. The exact wave function is referred to as anomalous. It is not a superposition of conventional eigenfunctions 
  and gets outside the Hilbert space. Anomalous electron and antielectron are possible. The antielectron is not a vacancy in the conventional valance band. The anomalous 
  electron-antielectron pair is created from the anomalous vacuum like electron-positron pair is created from the electron-positron vacuum. Formation of the anomalous vacuum 
  is not a single electron effect but the collective quantum phenomenon. In the film of graphen the anomalous states are located at the film edge and are expected to be of 
  high conductivity. 
  
  \end{abstract} \vskip 1.0cm



  \maketitle

  \section{INTRODUCTION}
  \label{intr}
  Dirac marvels are a broad class of matter \cite{MAT}. This class includes Dirac insulators ans semimetals, where electrons obey the Dirac like wave equation 
  \cite{PAN,KAN,WOL,FUS,XUI,LIU}. Graphene exhibits a lot of remarkable properties and belongs to two-dimensional Dirac systems \cite{GEI,PER}. Three-dimensional 
  Dirac systems, where there is the transition from the conventional insulator to topological one, are revealed \cite{XUI,LIU}. The topological aspects of Dirac materials 
  are revolutionary in condensed matter \cite{KAN}. 
  
  In conventional semiconductors and semimetals electrons are described by the Schr\"{o}dinger formalism. In Dirac materials like graphene electrons and holes are 
  interconnected obeying the Dirac equation instead of Schr\"{o}dinger one \cite{MAT,NIM,BERN}. In addition to this known issue, as shown in this paper, there is a 
  formation of the collective quantum state in graphene. Such state is absent in conventional semiconductors. 
  
  The solution of a stationary wave equation is a superposition of conventional eigenfunctions corresponding to vectors in the Hilbert space. But one can choose 
  expansion coefficients of such series in a special way. Namely, the series becomes convergent solely within the certain circle of the sample. (In this paper we
  consider two-dimensional Dirac materials like graphene, when each point $\{x,y\}$ is acribed the complex value $x+iy$.) Outside this region an analytic continuation is 
  required in the form of a different series. In this case, according to general rules, the exact wave function become singular. 
  
  We choose the square root singularity $\sqrt{x+iy-r_c}$. The singularity center has coordinates $\{{\rm Re}\,r_c, {\rm Im}\,r_c\}$. Thus, the wave function in an infinite 
  film  become double valued that is non-physical. However, if to choose the singularity position at the film edge, the $2\pi$-circulation around $r_c$ is impossible and 
  the wave function become single valued that is physical.  
  
  This state is referred to as anomalous. It looks as a spot on the film, which can be localized at any point of its edge. In the bulk Dirac material the spot is stretched 
  into a cylinder parallel to the sample surface and localized on it. The anomalous state is irreducible to conventional eigenfunctions. It does not correspond to a vector 
  in the Hilbertspace. 
  
  There are conventional electrons and holes in graphene. The hole is a vacancy in the valence band. Analogously, there are anomalous electron and antielectron. But the 
  antielectron is not a vacancy in the valence band. Like the positron is not a vacancy in the Dirac sea but can be created from the electron-positron vacuum. The same 
  way, the anomalous pair of electron and antielectron is created from the anomalous vacuum. The anomalous pair can be produced leaving the valance band completely filled. 
  
  The anomalous vacuum looks as the basic property of graphene. This is not a single particle effect but the  collective quantum state. This state appears during 
  formation of the graphene material.
  
  The edge states of the graphene film or the surface state in a three-dimensional sample are expected to be of large conductivity. 
  
  \section{DIRAC EQUATION IN TWO DIMENSIONS}
  \label{wave}
  In the relativistic quantum mechanics the wave function of a free electron obeys the stationary Dirac equation \cite{LANDAU2}
  \begin{equation}
  \label{1} 
  \left(\gamma^0\varepsilon_q+ic\pmb\gamma\cdot\nabla -\Delta\right)\psi(\pmb r)=0,
  \end{equation}
  where $\pmb r=(x,y,z)$ and $\varepsilon_q=\sqrt{q^2c^2+\Delta^2}>0$ is the electron spectrum. The bispinor $\psi$ and $\gamma$-matrices are
  \begin{equation}
  \label{2} 
  \psi=
  \begin{pmatrix}
  \Phi_1\\
  \Phi_2\\
  \Theta_1\\
  \Theta_2
  \end{pmatrix},
  \hspace{0.5cm}\pmb\gamma=
  \begin{pmatrix}
  0&\pmb\sigma\\
  -\pmb\sigma&0
  \end{pmatrix},
  \hspace{0.5cm}\gamma^0=
  \begin{pmatrix}
  1&0\\
  0&-1
  \end{pmatrix}.
  \end{equation}
  Solutions of the Dirac equation are analyzed, for example, in \cite{PAL,DIM}. The stationary solution is a superposition of conventional eigenfunctions 
  \cite{LANDAU2,LANDAU1} corresponding to vectors in the Hilbert space. 
  
  When the wave function does not depend on $z$, four coupled equations (\ref{1}) are separated on two independent pairs, $(\Phi_1,\Theta_2)$ and  $(\Phi_2,\Theta_1)$. 
  Below we consider the first pair with the notation $\Phi=\Phi_1$ and $\Theta=\Theta_2$ satisfying the equation 
  \begin{equation}
  \label{3} 
  \varepsilon_q
  \begin{pmatrix}
  \Phi\\
  \Theta
  \end{pmatrix}=\left(-ic\pmb\sigma\cdot\nabla+\sigma_z\Delta\right)
  \begin{pmatrix}
  \Phi\\
  \Theta
  \end{pmatrix},
  \end{equation}
  where $\nabla=(\partial/\partial x,\partial/\partial y)$. In components it reads 
  \begin{eqnarray}
  \label{4}
  (\varepsilon_q-\Delta)\Phi=-ic\left(\frac{\partial}{\partial x}-i\frac{\partial}{\partial y} \right)\Theta\\
  (\varepsilon_q+\Delta)\Theta=-ic\left(\frac{\partial}{\partial x}+i\frac{\partial}{\partial y} \right)\Phi
  \label{5}
  \end{eqnarray}
  
  The solution of (\ref{3}) is a superposition of various angular harmonics ($re^{i\varphi}=x+iy$) 
  \begin{equation}
  \label{18}
   \begin{pmatrix}
  \Phi\\
  \Theta
  \end{pmatrix}=\sum^{\infty}_{l=0}c_l
  \begin{pmatrix}
  \sqrt{\varepsilon_q+\Delta}\,J_l(qr)e^{il\varphi}\\
  i\sqrt{\varepsilon_q-\Delta}\,J_{l+1}(qr)e^{i(l+1)\varphi}
  \end{pmatrix},
  \end{equation}
  where $\varphi=\arctan y/x$ and the Bessel functions satisfy the relation $J_{-l}(z)=(-1)^lJ_l(z)$. At small argument $J_l(z)\sim z^{|l|}$. The total $\pmb r,t$
  dependent wave function includes the factor $e^{-it\varepsilon_q}$. 
  
  For a conventional wave function the expression (\ref{18}) holds in the entire space. In this case the wave function of the antiparticle, according to known rules
  \cite{LANDAU2}, has the form
  \begin{equation}
  \label{18a}
  \begin{pmatrix}
  \tilde\Phi\\
  \tilde\Theta
  \end{pmatrix}=\sum^{\infty}_{l=0}c_l
  \begin{pmatrix}
  \sqrt{\varepsilon_q-\Delta}\,J_{l-1}(qr)e^{i(l-1)\varphi}\\
  i\sqrt{\varepsilon_q+\Delta}\,J_l(qr)e^{il\varphi}
  \end{pmatrix}.
  \end{equation}
  The total $\pmb r,t$ dependent wave function includes the factor $e^{-it\varepsilon_q}$. 
  
  According to (\ref{18}) and (\ref{18a}), the total charge density is
  $e\langle(|\Phi|^2+|\Theta|^2-|\tilde\Phi|^2-|\tilde\Theta|^2)\rangle=e(\varepsilon_q-\Delta)\sum|c_l|^2[J^{2}_{l+1}(qr)-J^{2}_{l-1}(qr)]$. This difference of
  squared Bessel function is $-(2l/rq^2)\partial J^{2}_{l}(qr)/\partial r$ \cite{GRAD}. Thus, the total charge, after a pair creation is zero. 
  
  \subsection{Anomalous wave function}
  \label{who}
  In this section we study how the formation of the wave function can be followed step by step. The Bessel function at a fixed argument and $l\rightarrow+\infty$ is 
  \cite{GRAD}
  \begin{equation}
  J_l(z)=\frac{1}{\sqrt{2\pi l}}\left(\frac{ez}{2l}\right)^l=\frac{2l}{z}J_{l+1}(z).
  \label{19}
  \end{equation}
  Let us choose the expansion coefficients in (\ref{18}) in the form 
  \begin{equation}
  c_l=-\frac{1}{l\sqrt{2}}\left(\frac{2l}{eqr_c}\right)^l,\hspace{0.4cm}l\rightarrow+\infty
  \label{20}
  \end{equation}
  
  The large $l$ part of the series (\ref{18}) results in the singular contribution at $r<|r_c|$
  \begin{equation}
  \frac{\Phi}{\sqrt{\varepsilon_q+\Delta}}=-\frac{1}{2\sqrt{\pi}}\sum_l\frac{1}{l\sqrt{l}}\left(\frac{r}{r_c}e^{i\varphi}\right)^l\rightarrow \sqrt{1-\frac{x+iy}{r_c}}.
  \label{21}
  \end{equation}
  The limit of $l\rightarrow +\infty$ corresponds to the close vicinity $|x+iy-r_c|\ll |r_c|$ of the singularity point. According to (\ref{18}) and (\ref{19}), $\Theta$ has 
  the singular contribution at $r<|r_c|$
  \begin{equation}
  \Theta\rightarrow\frac{qr_c}{3}\sqrt{\varepsilon_q-\Delta}\left(1-\frac{x+iy}{r_c}\right)^{3/2}.
  \label{21a}
  \end{equation}
  
  The limit of large $l$ determines the radius of convergence. The series (\ref{21}) is convergent solely at $r<|r_c|$. Thus, under the choice (\ref{20}) of coefficients $c_l$, 
  the superposition (\ref{18}) of conventional eigenfunctions is not a solution of the Dirac equation at $r>|r_c|$.
  
  Close to the singularity at $x+iy=r_c$ large $l\sim 1/(x+iy-r_c)$ in (\ref{18}) are involved. Away from the singularity, even on the circle of convergence 
  $x^2+y^2=r^{2}_{c}-0$,  moderate  $l$ are essential. 
  
  The solution in the entire space can be obtained by the analytic continuation to $r>|r_c|$. The solution (\ref{18}) terminates at $r=|r_c|$ and the solution at $r>|r_c|$ is
  formed independently of (\ref{18}) just by matching it with the exact wave function at $r=|r_c|$. At $r>|r_c|$ the solution has the form 
  \begin{eqnarray}
  \nonumber
  &&\begin{pmatrix}
  \Phi\\
  \Theta
  \end{pmatrix}=\sum^{\infty}_{l=-\infty}(\tilde c_l+a_l  )
  \begin{pmatrix}
  \sqrt{\varepsilon_q+\Delta}\,N_l(qr)e^{il\varphi}\\
  i\sqrt{\varepsilon_q-\Delta}\,N_{l+1}(qr)e^{i(l+1)\varphi}
  \end{pmatrix}\\
  &&+\sum^{\infty}_{l=-\infty}b_l
  \begin{pmatrix}
  \sqrt{\varepsilon_q+\Delta}\,J_l(qr)e^{il\varphi}\\
  i\sqrt{\varepsilon_q-\Delta}\,J_{l+1}(qr)e^{i(l+1)\varphi}
  \end{pmatrix}.
   \label{22}
  \end{eqnarray}
  
  The coefficients $a_l$ and $b_l$ (\ref{22}) satisfy the conditions
  \begin{equation}
   |l||a_l|^{1/|l|}\bigg|_{|l|\rightarrow\infty}= \frac{|b_l|^{1/|l|}}{|l|}\bigg|_{|l|\rightarrow\infty}=0.
  \label{23}
  \end{equation}
  
  The Neumann function at a fixed argument and $l\rightarrow+\infty$ is \cite{GRAD}
  \begin{equation}
  N_l(z)=\sqrt{\frac{2}{\pi l}}\left(\frac{2l}{ez}\right)^l=\frac{2l}{z}N_{l-1}(z). 
  \label{24}
  \end{equation}
  Let us choose the expansion coefficients $\tilde c_l$ in (\ref{22}) in the form 
  \begin{equation}
  \tilde c_{l}=(-1)^{|l|+1}\frac{i}{2|l|\sqrt{2}}\left(\frac{eqr_c}{2|l|}\right)^{|l|}\theta(-l).
  \label{25}
  \end{equation}
  The large $|l|$ part of the series (\ref{22}) results in the singular contribution at $|r_c|<r$
  \begin{eqnarray}
  \nonumber
  &&\frac{\Phi}{\sqrt{\varepsilon_q+\Delta}}=-\frac{i}{2\sqrt{\pi}}\sum_{|l|}\frac{1}{|l|\sqrt{|l|}}\left(\frac{r_c}{r}e^{-i\varphi}\right)^{|l|}\\
  &&\rightarrow i\sqrt{1-\frac{r_c}{x+iy}}.
  \label{26}
  \end{eqnarray}
  The parts of (\ref{22}) with $a_l$ and $b_l$ do not contribute to the singularity due to the conditions (\ref{23}).
  
  According to (\ref{22}) and (\ref{24}), $\Theta$ has the singular contribution at $|r_c|<r$
  \begin{equation}
  \Theta\rightarrow-\frac{iqr_c}{3}\sqrt{\varepsilon_q-\Delta}\left(1-\frac{r_c}{x+iy}\right)^{3/2}.
  \label{27}
  \end{equation}
  
  The series (\ref{26}) is convergent at $r>|r_c|$. That is the radius of convergence in (\ref{21}) and (\ref{26}) is $r=|r_c|$ . According to general rules, the exact 
  wave function has singularities on the circle of convergence. In our case of (\ref{21}) and (\ref{26}) the singularity positions are determined by the condition 
  $e^{i\varphi}r/r_c=1$ equivalent to $x+iy=r_c$. 
  
  The exact wave function in the entire space has the form 
  \begin{equation}
  \label{28}
   \begin{pmatrix}
  \Phi\\
  \Theta
  \end{pmatrix}=
  \begin{pmatrix}
  \alpha(x,y)\sqrt{x+iy-r_c}+f(x,y)\\
  \beta(x,y)(x+iy-r_c)^{3/2}+g(x,y)
  \end{pmatrix},
  \end{equation}
  where $\alpha(x,y)$ and $\beta(x,y)$ at $x+iy=r_c$ follow from (\ref{26}) and (\ref{27}). The functions $\alpha$, $\beta$, $f$, and $g$ are non-singular. On the
  circle $x^2+y^2=|r_c|^2+0$ the smooth functions $f$ and $g$ are generated by $a_l$ and $b_l$ parts of (\ref{22}). 
  
  The non-singular parts $f(x,y)$ and $g(x,y)$ of the total solution at $x^2+y^2=|r_c|^{2}-0$ are determined by given $c_l$. Two sets $\{a_l\}$ and $\{b_l\}$ in 
  (\ref{22}) allow to match two functions $f(x,y)$ and $g(x,y)$ on the whole circle $x^2+y^2=r^{2}_{c}+0$ to analogous ones at $x^2+y^2=|r_c|^2-0$. This is similar to 
  matching of a function and its normal derivative in the usual boundary problem. 
  
  The analytic continuation of the wave function to the entire space of $x+iy$ produces the double valued function. Such wave function is not physical. However, when 
  this solution is localized at the border of a sample, occupying the region $y>0$, the analytic continuation through the region $y<0$ is impossible since $y$ cannot 
  be negative. Under these conditions the resulting wave function becomes single valued (Sec.~\ref{mat}). This situation is studied below. 
  
  \subsection{Comments}
  \label{com}
  \begin{itemize}
  \item
  The wave function, obtained in Sec.~\ref{who}, is referred to as anomalous. The key attribute of the anomalous function is that it is not a superposition of
  conventional eigenfunctions. Such superposition exists not in the entire space but just withing the circle of convergence. In this region each angular harmonics 
  can be re-expanded in Dirac plane waves \cite{LANDAU2,AKH}. Thus the total anomalous wave function at $x^2+y^2<|r_c|^2$ looks as a superposition of Dirac plane waves.
  But this series is divergent. 
  
  However, the exact anomalous wave function in the entire space can be expanded in Fourier harmonics. These harmonics contain contributions with momenta 
  $|\pmb p|\sim 1/|r_c|$ but they are not of the type $\delta(|\pmb p|-q)$ as it should be for conventional Dirac plane waves, which are on the mass surface 
  \cite{LANDAU2,AKH}. Thus, the anomalous particle is a non-Dirac fermion.
  
  \item
  It could be various types of singularities at various points in the complex plane. It is not clear what are ``quantum numbers'' since, even for a fixed type of 
  singularity, each given set $\{c_l\}$ generates a particular shape of the anomalous wave function. Different shapes correspond to different ``eigenfunctions''.  
  Anomalous wave functions cannot be classified as vectors in the Hilbert space.
  
  \item
  Electrons and holes in graphene are interconnected obeying the Dirac wave equation but not Schr\"{o}dinger one. Accordingly, the anomalous wave function can be
  collected either from electron like or hole like ones. The former and latter are referred to as anomalous electron and anomalous antielectron. The property of pair
  creation is attributed to a Dirac material. But the antielectron is not a vacancy in the valence band like the positron is not a vacancy in the Dirac sea. Instead the 
  electron-positron pair is created from the electron-positron vacuum. Analogously, the certain anomalous vacuum should exist, where the anomalous pair is created from. 
  
  Note that in a conventional semiconductor electrons and holes are not interconnected and obey the Schr\"{o}dinger equation (Sec.~\ref{ato}). Anomalous electron and 
  antielectron can exist in these materials as individual particles (Sec.~\ref{ato}), which are not associated with the anomalous vacuum.
  
  \end{itemize}
   
  \section{ANOMALOUS STATES IN GRAPHENE}
  \label{mat}
  Dirac materials are associated with the electron Dirac spectrum \cite{MAT}. Graphene is a two dimensional example of Dirac materials consisting of two sublattices $A$ and
  $B$ \cite{CAS}. The tight binding approximation is appropriate for a description of the electron spectrum \cite{MAT}
  \begin{equation}
  \label{29}
  \hat H=-t\sum_{i,j}\left[a^+(\pmb r_i)b(\pmb r_J)+b^+(\pmb r_j)a(\pmb r_i)\right],
  \end{equation}
  where $\pmb r_{i}$ referrers to the sublattice $A$ and $\pmb r_j=\pmb r_{i}+\pmb\delta$ referrers to the sublattice $B$, where
  $\pmb\delta=(\pmb\delta_1,\pmb\delta_2,\pmb\delta_3)$ accounts for three nearest neighbors \cite{CAS}. The parameter $t\simeq 2.7\,eV$. There are two Dirac points, 
  $\pmb K$ and $\pmb K'$, in the Brillouin zone, where the electron spectrum is conical. Below we consider a vicinity of the $\pmb K$ point and use the definition 
  \begin{equation}
  \label{30}
  \begin{pmatrix}
  \Phi(\pmb r_i)\\
  \Theta(\pmb r_j)
  \end{pmatrix}=
  \begin{pmatrix}
  a(\pmb r_i)\exp(-i\pmb K\cdot\pmb r_i)\\
  b(\pmb r_j)\exp(-i\pmb K\cdot\pmb r_j)
  \end{pmatrix} 
  \end{equation}
  corresponding to pseudospin two due to two sublattices. (In the case of $\pmb K'$ point one has to change $\pmb K\rightarrow \pmb K'$). 
  
  According to the tight binding formalism, the functions $\Phi$ (sublattice $A$) and $\Theta$ (sublattice $B$) modulate the row of usual atomic wave functions 
  $\psi_{A,B}$. The electron wave function is
  \begin{eqnarray}
  \nonumber
  &&\sum_i\Phi(\pmb r_i)\psi_{A}(\pmb r-\pmb r_i)\exp(i\pmb K\cdot\pmb r_i)\\
  &&+\sum_j\Theta(\pmb r_j)\psi_{B}(\pmb r-\pmb r_j)\exp(i\pmb K\cdot\pmb r_j)
  \label{31}
  \end{eqnarray}
  with the envelope functions $\Phi(\pmb r_i)$ and $\Theta(\pmb r_j)$. The transformations
  \begin{eqnarray}
  \nonumber
  &&\Phi(\pmb r_i)=\frac{1}{\sqrt{N/2}}\sum_{\pmb q}e^{i\pmb q\cdot\pmb r_i}\Phi_{\pmb q}\\
  &&\Theta(\pmb r_j)=\frac{1}{\sqrt{N/2}}\sum_{\pmb q}e^{i\pmb q\cdot\pmb r_J}\Theta_{\pmb q},
  \label{32}
  \end{eqnarray}
  contain Wannier exponents. Here $N/2$ is the number of $A(B)$ sites. The modified Hamiltonian (\ref{29}) is
  \begin{equation}
  \label{33} 
  \hat H=\sum_{\pmb q}
  \begin{pmatrix}
  \Phi^{+}_{\pmb q}\\
  \Theta^{+}_{\pmb q}
  \end{pmatrix}
  \begin{pmatrix}
  0&\xi_{\pmb q}\\
  \xi^{*}_{\pmb q}&0
  \end{pmatrix}
  \begin{pmatrix}
  \Phi_{\pmb q}\\
  \Theta_{\pmb q}
  \end{pmatrix},
  \end{equation}
  where at $q\ll 1/a$ ($a$ is the unit size of the crystal lattice) $\xi_{\pmb q}=c(q_x-iq_y)$ \cite{MAT}.
  
  With the transformation inverse to (\ref{32})
  \begin{eqnarray}
  \nonumber
  &&\Phi_{\pmb q}=\frac{1}{\sqrt{N/2}}\sum_{i}e^{-i\pmb q\cdot\pmb r_i}\Phi(\pmb r_i)\\
  &&\Theta_{\pmb q}=\frac{1}{\sqrt{N/2}}\sum_{j}e^{-i\pmb q\cdot\pmb r_j}\Theta(\pmb r_j),
  \label{34}
  \end{eqnarray}
  the Hamiltonian (\ref{33}) takes the form 
  \begin{equation}
  \label{35} 
  \hat H=\sum_{i,j}
  \begin{pmatrix}
  \Phi^{+}(\pmb r_i)\\
  \Theta^{+}(\pmb r_j)
  \end{pmatrix}
  \begin{pmatrix}
  0&F(\pmb r_i-\pmb r_j)\\
  F^{*}(\pmb r_i-\pmb r_j) &0
  \end{pmatrix}
  \begin{pmatrix}
  \Phi(\pmb r_i)\\
  \Theta(\pmb r_j)
  \end{pmatrix}
  \end{equation}
  where
  \begin{equation}
  \label{36}
  F(\pmb r_i-\pmb r_j)=\frac{1}{N/2}\sum_{\pmb q}\xi_{\pmb q}\,e^{i\pmb q\cdot(\pmb r_i-\pmb r_j)}.
  \end{equation}
  
  Suppose now that the fractions of the total sets $\{\Phi(\pmb r_i)\}$ and $\{\Theta(\pmb r_j)\}$ are slow varying functions compared to the inter-atomic distance $a$.
  Then one can use for these fractions the continuous approach in (\ref{35}). In this limit of large distance only small $q\ll 1/a$ are essential. With the relation 
  \begin{equation}
  \label{37}
  \sum_{\pmb q}e^{i\pmb q\cdot(\pmb r_i-\pmb r_j)}=\delta(\pmb r_i,\pmb r_j)N/2
  \end{equation}
  the Hamiltonian (\ref{35}) acquires the form
  \begin{eqnarray}
  \label{38} 
  &&\hat H=\int d^3r
  \begin{pmatrix}
  \Phi^{+}(\pmb r)\\
  \Theta^{+}(\pmb r)
  \end{pmatrix}\\
  &&\cdot\begin{pmatrix}
  0&c(-i\partial/\partial x-\partial/\partial y)\\
  c(-i\partial/\partial x+\partial/\partial y)&0
  \end{pmatrix}
  \begin{pmatrix}
  \Phi(\pmb r)\\
  \Theta(\pmb r)
  \end{pmatrix},
  \nonumber
  \end{eqnarray}
  which corresponds to the wave equation (\ref{3}) with $\Delta=0$. Thus, anomalous states there are also in graphene. A representation of proper wave functions as 
  combinations of Wannier exponents (\ref{32}) is impossible since these series are divergent (Sec.~\ref{com}).
  
  \subsection{Motion around the singularity point}
  \label{poi}
  The anomalous solution in graphine (Sec.~\ref{who}) is double valued corresponding to the branch point (\ref{28})
  \begin{equation}
  \label{40a}
  \sqrt{x+iy-r_c}.
  \end{equation}
  This is the continuous limit $\Phi(\pmb r_i)\rightarrow\Phi(\pmb r)$ and $\Theta(\pmb r_j)\rightarrow\Theta(\pmb r)$ of large distance, when one can ignore discreteness 
  of the crystal lattice. The double valued wave function is non-physical. It is obvious that this function cannot be expanded in conventional eigenfunctions. 
  
  The resulting function acquires a different value after the $2\pi$-circulation around the singularity point. This general property does not depend on that were we are: far 
  from the singularity point or close to it. On the short distance, where the lattice discreteness is essential, each lattice site $\pmb r_i$ is assigned the complex 
  number $r_{ix}+ir_{iy}$. The circulation around the singularity point occurs by jumps from one lattice site to another. The singularity point, localized at the certain 
  lattice site $\pmb r_c$, results in the singularity
  \begin{equation}
  \label{41}
  \sqrt{r_{ix}-r_{cx}+i(r_{iy}-r_{cy})}
  \end{equation}
  instead of the continuous limit (\ref{40a}).
  
  When the lattice site $\pmb r_c$ is placed exactly on the border of the sample, the circulation around this point is impossible and the wave function becomes single 
  valued that is physical. On the language of the continuous limit (\ref{40a}) this means that $y$ cannot be negative excluding thus a possibility of $2\pi$-circulation. 
  Those wave functions are referred to as anomalous. 
  
  We see that anomalous states can exist in graphene in the form of anomalous envelopes $\Phi(\pmb r_i)$ and $\Theta(\pmb r_j)$ in the wave function (\ref{31}). The states, 
  when these functions are slow varying, correspond to the continuous limit describing by Eq.~(\ref{3}). Note that Dirac matrices in this equation relate to pseudospin formed 
  by two sublatices \cite{MAT}. Since the spin-orbit interaction is absent, equations (\ref{3}) correspond either to spin up or spin down.
  
  \subsection{Two types of states in graphene}
  \label{two}
  It follows that two different classes of states are associated with graphene. 
  
\begin{figure}
  \includegraphics[width=6cm]{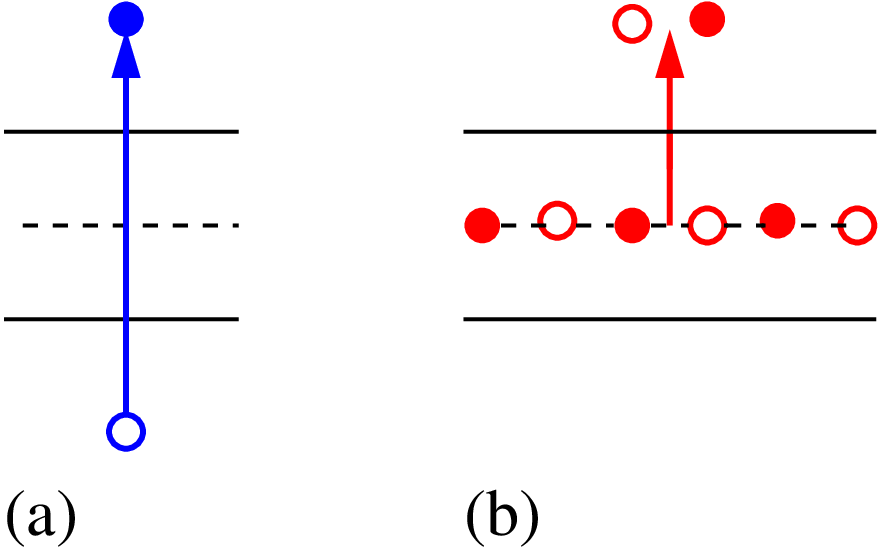}
  \caption{\label{fig1} (a) Creation of the conventional electron-hole pair leaves a vacancy in the valence band. (b) Creation of the anomalous electron-antielectron pair 
  occurs from the anomalous vacuum.}
  \end{figure}

  First, this is conventional electrons of conduction and valence bands. Each particle is ascribed a few quantum numbers sweeping various values. For example, for a free
  particle this is $q,j,l,m$ \cite{LANDAU2}. Creation of the electron-hole pair is a conventional process marked by the arrow in Fig.~\ref{fig1}(a).
  
  Second, this is anomalous states. In a thin film they are like spots, which can be localized at any point of the film edge. In the bulk Dirac material the spot is stretched 
  into a cylinder parallel to the sample surface and localized on it. 
  
  Besides the anomalous electron there is also its antiparticle referred to as antelectron. A pair of these particles can be created from the different vacuum called 
  anomalous (Sec.~\ref{com}). This is the process in Fig.~\ref{fig1}(b). The antielectron is a positively charged quasiparticle, which is not a vacancy in the valence 
  band as shown in Fig.~\ref{fig1}(b). The pair of anomalous electron and antielectron can be created from the anomalous vacuum keeping the valence band completely filled. 
  
  Analogously, the positron is not a vacancy in the Dirac sea. The electron-positron pair originates from the electron-positron vacuum similar to Fig.~\ref{fig1}(b). 
  
  In contrast to the conventional particle, the anomalous one depends on the continuous function of coordinates (Sec.~\ref{com}). A variation of this function at each point
  results in the different anomalous particle. Thus, there is an infinite set of ``quantum numbers'' related to points in the space. This results in a high density of states. 
  Accordingly, a high conductivity of edge electrons (surface electrons in a 3D sample) is expected. 
  
  Despite the delicate origin of anomalous particles they are expected to be thermally distributed above the gap along with conventional particlesin Fig.~\ref{fig1}.
  
  \section{ANOMALOUS STATES IN NON-DIRAC MATERIALS}
  \label{ato}
  The two-dimensional electron system is realized in atomically thin films fabricated and studied in \cite{GRI}. The films are stable and exhibit a long rang ballistic 
  transport of electrons. There is no evidence that those films belong to Dirac materials due to the experimental indication of the Schr\"{o}dinger electron spectrum 
  $(q^{2}_{x}+q^{2}_{y})/2m$ at small $q$. Another branch, with the opposite sign of mass, was shifted in momentum. 
  
  In the situation of two-dimensional Schr\"{o}dinger spectrum one component wave function $\Phi$, satisfying the Schr\"{o}dinger equation in two dimensions, is equivalently 
  described by Eqs. (\ref{3})
  \begin{eqnarray}
  \nonumber
  &&-\frac{1}{2m}\left(\frac{\partial}{\partial x}-i\frac{\partial}{\partial y}\right)\Theta=E\Phi\\
  &&\Theta=\left(\frac{\partial}{\partial x}+i\frac{\partial}{\partial y}\right)\Phi
  \label{39}
  \end{eqnarray}
  if to put $2mc^2E=\varepsilon^{2}_{q}-\Delta^2$. Thus, anomalous states also exist in the Schr\"{o}dinger system in two dimensions. So that anomalous states are not 
  attributed to Dirac materials only.
  
  The measurements, reported in \cite{GRI}, related to conventional electrons and holes since the anomalous component was absent in those non-Dirac materials. 
  
  Besides the electron spectra there is another essential difference between Dirac and non-Dirac materials. In conventional semiconductors anomalous electron and 
  antielectron exist as individual particles. Creation of them from the anomalous vacuum is impossible since this vacuum exists solely in Dirac materials. Thus, anomalous 
  states in non-Dirac materials have just a theoretical aspect.  
  
  \section{DISCUSSIONS}
  \label{disc}
  A stationary solution of a quantum mechanical wave equation is a superposition of eigenfunctions. Each eigenfunction corresponds to a vector in the Hilbert space. 
  
  But graphene is a ``magic'' matter, where electrons and holes are interconnected resulting in the Dirac states instead of Schr\"{o}dinger ones. This is not a whole 
  story. There is another basic property of graphene: this material is characterized by the collective quantum state. This state appears as a formation of the anomalous 
  vacuum, which is not a single particle effect but the collective phenomenon. The anomalous vacuum is a source of anomalous electron-antielectron pairs. States of these 
  particles do not correspond to vectors in the Hilbert space. 
  
  Due to the high density of anomalous states the high conductivity of edge electrons (surface electrons in a 3D sample) is expected. The anomalous states contribute to 
  the total conductivity of a sample together with the high conductivity in bulk because of small effective mass in graphene.
  
  Anomalous pairs are thermally generated from the anomalous vacuum in graphene. In conventional semiconductors anomalous particle can also exist but they cannot be 
  created from the anomalous vacuum because it does not exist in non-Dirac materials. Thus in metals or conventional semiconductors anomalous states have just a 
  theoretical aspect.
  
  Besides graphens the anomalous states can be also formed in $d$-wave superconductors. The anomalous formalism is applicable to the massless magnetic monopole 
  mediated by the chiral field \cite{LOSH}.
  
  \section{CONCLUSIONS}
  \label{conc}
  The electron state of the graphene sample has more basic aspects than the Dirac spectrum only. This state is the collective quantum one characterized by the anomalous 
  vacuum. This is not a single electron effect but the collective phenomenon. Anomalous electron-antielectron pairs can be created from that vacuum like electrons and 
  positrons. States of anomalous particles do not correspond to vectors in the Hilbert space. 
  
  \acknowledgments

  I am grateful to J. Engelfried for discussions of related topics. This work was supported by CONACYT through grant 237439.

  \end{document}